\documentclass[twocolumn]{revtex4-1}
\usepackage{graphicx}
\usepackage{color}
\newcommand{\ket}[1]{\left| #1 \right\rangle}

\newcommand{\proj}[1]{| #1\rangle\!\langle #1 |}
\newcommand{\Tr}{\mathrm{Tr}}

\newcommand{\eea}{\end{eqnarray}}
\newcommand{\bea}{\begin{eqnarray}}
\newcommand{\ee}{\end{equation}}
\newcommand{\be}{\begin{equation}}

\begin{document}
\title{Self-consistent tomography and measurement-device independent cryptography}

\author{I.D. Moore and S.J. van Enk}

\affiliation{Department of Physics and
Oregon Center for Optical, Molecular \& Quantum Sciences\\
University of Oregon, Eugene, OR 97403}

\begin{abstract}

A recurring problem in quantum mechanics is to estimate either the state of a quantum system or the measurement operator applied to it. If we wish to estimate {\em both}, then the difficulty is that the state and the measurement always appear together: to estimate the state, we must use a measurement; to estimate the measurement operator, we must use a state. The data of such quantum estimation experiments come in the form of measurement frequencies. Ideally, the measured average frequencies can be attributed to an average state and an average measurement operator. If this is not the case, we have correlated state-preparation-and-measurement (SPAM) errors. We extend some tests developed to detect such correlated errors to apply to a cryptographic scenario in which two parties trust their individual states but not the measurement performed on 
the joint state.

\end{abstract}

\maketitle
\section{Introduction}

Quantum-state tomography started from the question whether a quantum-mechanical wave function is a measurable thing. The answer is clear now, a quantum state can be reconstructed from data from a sufficient number of different measurements on many systems prepared in the same state \cite{fano1957,paris2010}.
One condition for this reconstruction to work 
is that the state-preparation procedure is reproducible, i.e., that it indeed produces a well-defined state, at least on average (there may well be (small) random fluctuations). 
A second condition is that we know what measurements we actually perform.

In recent years we have come to realize that the latter condition is not so easy to fulfill, and that often we have just as much or as little control about the states we prepare as about the measurements we do. That is, SPAM (state-preparation and measurement) errors are inevitable.
In response, self-consistent versions of tomography have been introduced that aim to estimate both quantum states {\em and} measurements [and even quantum gates, applied after state preparation and before measurement, as well] in a self-consistent manner, such that the measurement data are fully explained by 
those state and measurement descriptions \cite{mogil2012,branczyk2012,merkel2013,blume2013,medford2013,stark2014,greenbaum2015}.
These schemes have become important for diagnosing small and subtle errors in small quantum computing devices.

One can go one step further than self-consistent tomography and perform overcomplete sets of measurements on overcomplete sets of states that allow one to check the assumption that a given state-preparation procedure indeed produces a single specific state, and that a given measurement procedure indeed produces a single specific measurement.
In particular, ``holonomic'' SPAM tomography or ``loop'' SPAM tomography denote a procedure to check for correlations between measurement and state-preparation \cite{jackson2015,beck2017}. For example, suppose one uses a laser to perform a measurement on a system whose state was prepared using that same laser just a microsecond ago. It may well be that the measurement performed depends on (and so is correlated with) what state was created. Importantly, checking for this type of correlations with the protocol from Ref.~\cite{jackson2015} does {\em not} require one to either reconstruct or know the measurement and state operators.

With multiple quantum systems there may be correlations between different measurements or between different state preparations as well. Those correlations, too, can be detected (again, without the need for reconstructing state and measurement operators) \cite{jackson2017,jackson2017b,beck2018}.
We focus here on the case of two qubits A and B and as usual we assume these are controlled by Alice and by Bob, respectively.
There are two scenarios of interest to us:  (i) Alice and Bob share an entangled state and perform separate measurements, each on their own qubit, (ii) Alice and Bob separately prepare states of their own qubit and then send their qubits to a device that performs a joint measurement. The probability to get a particular measurement outcome can be written for both scenarios in the form
\be\label{P}
P=\Tr (\alpha_A\otimes \beta_B\xi_{AB} ).
\ee
Here, in scenario (i), $\xi$ denotes the joint state of qubits A and B, and $\alpha$ and $\beta$ represent the single-qubit measurements Alice and Bob perform, respectively. In scenario (ii) $\xi$ denotes the joint measurement and  $\alpha$ and $\beta$ represent the single-qubit states Alice and Bob prepare, respectively. Mathematically these two scenarios are obviously very similar: one difference is that measurement operators are not normalized while states (represented by density operators) are. In both scenarios the idea is for Alice and Bob to gain information about $\xi$ by trying out different (state or measurement) operators $\alpha_A^{(k)}$ and $\beta_B^{(l)}$. Here $k,l=1\ldots N$, where $N$ is some suitable number large enough to be able to tomographically reconstruct $\xi$ {\em if} the operators $\alpha_A^{(k)}$ and $\beta_B^{(l)}$ were all known (for more details, see the next Section).

Scenario (i) was implemented experimentally recently \cite{beck2018}. Even if $\xi$ is a product state, a Bell inequality may be violated [a ``fake violation''] if Alice's and Bob's measurements are sufficiently strongly correlated.
Loop tomography allows Alice (and by symmetry, Bob, too) to check for exactly those correlations.

Here we're interested in the second scenario, 
whose setting is as follows. Alice and Bob each prepare individual qubits that they then send to a measurement device ({\em not} under their control) that is supposed to perform a particular joint two-outcome measurement (e.g., projecting onto a singlet state such that $\xi=\proj{\rm singlet}$: the two outcomes are ``yes'' or ``click'' and ``no'' or ``no click'', the latter described by the operator $\openone-\xi$).
The device may, however, perform a different measurement that, in addition, may depend on what states Alice and Bob sent. For example, an eavesdropper may have some (imperfect) knowledge of which state either Alice or Bob sent and adjust the measurement. The measurement procedure is then described by an operator $\xi$ that depends on either Alice's or Bob's input states. We will show how Alice and Bob can exploit  loop tomography to detect that type of measurement dependence on their own states. 

In spite of the great mathematical similarity between the two scenarios, there is one crucial difference.
In scenario (ii) the measurement $\xi$ produces just one bit of information, as opposed to the two bits of information produced by Alice's and Bob's single-qubit measurements together in scenario (i). This will mean that Alice will have to make use of her knowledge of her state-preparation procedure to find correlations between $\xi$ and her states. (Similarly for Bob.)
In the experimental implementation of scenario (i) \cite{beck2018} no such knowledge was necessary. 

Note that scenario (ii) is exactly the setting of measurement-device independent quantum cryptography (including the assumption Alice and Bob do know their own states, but not the measurement $\xi$)  \cite{lo2012}. Similarly, the recently developed twin-field quantum cryptographic protocol \cite{twin2018,twin2019a,twin2019b} fits our scenario (ii), and so does the qubit version of that protocol \cite{twin2019c}. 
On the other hand, Ekert's quantum key distribution (QKD) protocol uses separate measurements and jointly prepared states \cite{ekert1991}, and thus fits scenario (i).

It is now high time to provide a fully detailed description of our main scenario (ii) and prove our assertions. 
\section{Loop SPAM tomography}
\subsection{Scenario (ii)}
If Alice and Bob together would like to be able to tomographically reconstruct the two-qubit measurement operator $\xi$, they each need to prepare $4$ different (linearly independent) states of their qubits. 
Given probabilities of the form (\ref{P}), Alice and Bob can gather the measured frequencies of detector ``clicks'' 
in a $4$-by-$4$ data matrix
whose expectation value should ({\em if} there is a unique single-valued operator $\xi$) have the form
\be\label{Fkl}
F^{kl} = \textrm{Tr}\{\alpha_A^{k}\otimes\beta_B^{l}\, \xi_{AB}\}
\ee
for $k,l=1\ldots 4$.
The trace on the right-hand-side of Eq.~(\ref{Fkl}) can be calculated by expanding all operators in the Pauli basis as follows
\be
 \alpha^{k} = \sum_{i} \alpha_{i}^{k}\, \sigma_{i}, \:\:\:\:\: \beta^{l} = \sum_{i} \beta_{i}^{l}\, \sigma_{i}, \:\:\:\:\: \xi = \sum_{i,j}x_{ij}\, \sigma_{i}\otimes\sigma_{j}.\ee
\noindent Substituting these expansions into the definition of the data matrix and noting that 
\be\textrm{Tr}\{\sigma_{i}\sigma_{j}\otimes\sigma_{k}\sigma_{l}\} = 4 \delta_{ij}\delta_{kl}, 
\ee
yields the equation
\be
F^{kl} =4 \sum_{i,j} x_{ij} \alpha^{k} _{i}\beta^{l}_{j}.
\ee
Eliminating the factor of four by defining $S=F/4$,  we can rewrite the equation for $S$ as a matrix equation
\be
S = A^{T} X B
\ee
where $X$ has matrix element $x_{ij}$ and $A$ has columns made from vectors of the Pauli expansion coefficients of Alice's operators, and $B$ is similarly defined for Bob. Multiplying both sides of this equation on the left by $(A^{T})^{-1}$ [we assume the inverse exists, i.e.,  we assume Alice's 4 operators to be linearly independent] yields
\be
(A^{T})^{-1} S = X B.
\ee
Now we assume that both Alice and Bob have control over their own operators such that these operators do not vary over the course of the experiment. Next, suppose that there are two trials, each using a different (not identical) set of operators $\alpha^k$ for Alice but the same set of operators $\beta^l$ for Bob. Then we can eliminate the unknown matrix $X$ and write the condition on there being a unique $X$ that does not depend on which operators $\alpha^k$ Alice is using, as
\be
(A_{1}^{T})^{-1} S_{1} = (A_{2}^{T})^{-1} S_{2},
\ee
where the left-hand and right-hand sides of the equation represent trials with different operators $\alpha^{k'}$.
One alternative useful way of rewriting this same equation is
\be
S_{1}S_{2}^{-1}A_{2}^{T}(A_{1}^{T})^{-1} = \openone,
\ee
even though this may fail, namely, if $S_2$ is not invertible.
It is helpful to write this matrix product out in terms of the coefficients:
\be\label{testA}
\sum_{k,l,m}(S_{1})_{ik} (S_{2}^{-1})_{kl} (A_{2}^{T})_{lm} ((A_{1}^{T})^{-1})_{mj}  = \delta_{ij}
\ee
Written this way, it is easy to see how this equation can be used to test for dependence of $\xi$  on Alice's operators: suppose, without loss of generality, that $\xi$ is somehow different for the operator $\alpha_A^i$ of the first trial. Then every row but  the $i$th row of $S_{1}$ is inverted by the remaining three matrices. Therefore, only the $i$th row of the left-hand side matrix will differ from the identity. Conversely, since Alice is able to calculate the left-hand side of (\ref{testA}) just from her knowledge of her operators $\alpha^k$ and from the data matrices $S_1$ and $S_2$ she can diagnose, without needing any knowledge about Bob's operators $\beta^l$, with which of her state preparations the measurement is  correlated. She also does not need to reconstruct the measurement operator $\xi$.

Note that all we need for this to work is that the trials are not identical. That is, Alice needs just 5 different states at a minimum [to make two non-identical sets of 4 states] to be able to run this check.

(By symmetry, Bob could diagnose the presence of correlations between $\xi$
and {\em his} states $\beta^l$ without needing knowledge of Alice's operators. In this case Bob would need to prepare at least 5 different states.)

\subsection{Example}
Consider the following example, where, for simplicity, we will ignore statistical fluctuations in the observed frequencies.

In each run,
Bob prepares one of  four different states $\ket{\psi_k}$ for $k=1\ldots 4$,
\bea
\ket{\psi_1}&=&\frac{\ket{0}+\ket{1}}{\sqrt{2}}\nonumber\\
\ket{\psi_2}&=&\frac{\ket{0}+i\ket{1}}{\sqrt{2}}\nonumber\\
\ket{\psi_3}&=&\frac{\ket{0}-\ket{1}}{\sqrt{2}}\nonumber\\
\ket{\psi_4}&=&\ket{0}
\eea
Bob will tell Alice what number $k$ he picked, but does not reveal what the corresponding states are.

Alice chooses five different states, the same four states that Bob chooses from, and in addition 
\be
\ket{\psi_5}=\ket{1}.
\ee
Both Alice and Bob send their qubits to a measurement device located somewhere between their labs. The device projects onto the singlet state, except when Alice sends a qubit in state number 5. In that case, the measurement projects onto, say, the symmetric state
\be
\ket{{\rm symm}}=\frac{\ket{00}+\ket{11}}{\sqrt{2}}.
\ee
If Alice gathers the relative frequencies of the ``yes'' outcomes for her states $k=1\ldots 4$ in combination with Bob's choices $k'$ in the matrix $S_1$ (as defined above) she finds (ignoring statistical fluctuations!)
\bea
S_1=
\left(
\begin{array}{cccc}
 0 & 1/4  & 1/2 & 1/4\\
 1/4 & 0  & 1/4  & 1/4\\
  1/2& 1/4  & 0  & 1/4\\
 1/4 &1/4 &1/4 & 0
\end{array}
\right).
\eea
This corresponds to the device projecting onto the singlet state. For example, on the diagonal we find 0's because the input state is symmetric;
the entries for $k=1,k'=3$ and $k=3,k'=1$ are 1/2, because the input is a product state in the antisymmetric subspace.
Her matrix $A_1$ describing the expansion coefficients of her first 4 states in the Pauli basis is
\bea
A_1=
\left(
\begin{array}{cccc}
 1/2 & 1/2  & 1/2 & 1/2\\
 1/2 & 0  & -1/2  & 0\\
  0& -1/2  & 0  & 0\\
 0 &0 &0 & 1/2
\end{array}
\right).
\eea
Multiplying $(A_1^T)^{-1}$ by $S_1$ gives her the result
\bea\label{AS1}
(A_1^T)^{-1} S_1=
\left(
\begin{array}{cccc}
 1/2 & 1/2  & 1/2 & 1/2\\
 -1/2 & 0  & 1/2  & 0\\
  0& 1/2  & 0  & 0\\
 0 &0 &0 & -1/2
\end{array}
\right).
\eea

If Alice includes the results from her fifth choice by replacing her fourth state choice, then she finds
\bea
S_2=
\left(
\begin{array}{cccc}
 0 & 1/4  & 1/2 & 1/4\\
 1/4 & 0  & 1/4  & 1/4\\
  1/2& 1/4  & 0  & 1/4\\
 1/4 &1/4 &1/4 & 0
\end{array}
\right).
\eea
The first three rows here are the same as in $S_1$ because they describe exactly the same data.
The fourth row pertains to her using the state $\ket{\psi_5}=\ket{1}$ instead of $\ket{\psi_4}=\ket{0}$, and, unbeknownst to her, that row corresponds to the different measurement $\proj{{\rm sym}}$.
Her state matrix $A_2$ is now
\bea
A_2=
\left(
\begin{array}{cccc}
 1/2 & 1/2  & 1/2 & 1/2\\
 1/2 & 0  & -1/2  & 0\\
  0& -1/2  & 0  & 0\\
 0 &0 &0 & -1/2
\end{array}
\right).
\eea
where the last column is different from that in $A_1$.

Multiplying $(A_2^T)^{-1}$ with $S_2$ now gives her
\bea
(A_2^T)^{-1} S_2=
\left(
\begin{array}{cccc}
 1/2 & 1/2  & 1/2 & 1/2\\
 -1/2 & 0  & 1/2  & 0\\
  0& 1/2  & 0  & 0\\
 0 &0 &0 & 1/2
\end{array}
\right).
\eea
This should be equal to the matrix in (\ref{AS1}), but
Alice notices the discrepancy in the last row, and concludes (correctly) that the measurement performed must have been different for states $\ket{\psi_4}$ and $\ket{\psi_5}$.

Alternatively, calculating the single matrix
$S_{1}S_{2}^{-1}A_{2}^{T}(A_{1}^{T})^{-1}$---which works here because $S_2$ is indeed invertible---gives her the result
\be
S_{1}S_{2}^{-1}A_{2}^{T}(A_{1}^{T})^{-1}=
\left(\begin{array}{cccc}
 1 & 0  & 0 & 0\\
 0 & 1  & 0  & 0\\
  0& 0  & 1  & 0\\
 1 &0 &1 & 1
\end{array}
\right).
\ee
Again, the last row is not what it should be for the right-hand side to be the identity matrix, leading Alice to the same correct conclusion that there is a correlation between the measurement performed and either state number 4 or state number 5. 

Of course, by subsequently replacing, say, state number 1 by state number 5, the same calculation would show her that either state 5 or state 1 is to blame. (Note she does not have to do a new experiment, just construct new matrices from the same data.)
Combining the two results then conclusively points to a correlation between her state number 5 and the joint measurement.
\subsection{Application to quantum cryptography}
As mentioned in the Introduction, scenario (ii)
fits particular cryptographic protocols very well, measurement-device independent cryptography \cite{lo2012}, as well as the improved ``twin-field'' versions \cite{twin2018,twin2019a,twin2019b,twin2019c}.
In these protocols, Alice and Bob each repeatedly and independently prepare qubits in certain (secret) quantum states, chosen from a fixed set, and send them  to a measurement device, which is located halfway between their labs. This device is supposed to do a particular measurement (e.g., projecting onto the singlet state), but since it is not under Alice's or Bob's control, they must verify the measurement results
for a randomly chosen small subset $R$. For example, Bob could communicate to Alice what states he prepared (just for the set $R$). Alice can verify, for example, that in all those cases where Bob sent the same state as she sent, the singlet outcome never occurred.

Note that in our protocol there is no need for Bob to tell Alice what state he sent, which is an advantage. On the other hand, in our protocol Alice and Bob will have to choose from a larger set of possible states (containing at least 5 different states) than in Ref.~\cite{lo2012}'s protocol. We can interpret  the additional states as ``decoy states.''
The idea of decoy states---which do not contribute to generating a secret bit but are meant to better detect an eavesdropper---has been known to be very useful in cryptographic contexts \cite{lo2005}. 
 Thus, we suggest that our protocol can be smoothly incorporated into the standard measurement-device independent cryptography protocol as well as into the recently developed twin-field versions \cite{twin2018,twin2019a,twin2019b,twin2019c} of that protocol.

\section{Conclusion}
In conclusion, we presented a tomographic protocol that is meant to detect correlations between measurements and state preparation, without the need to know or estimate what the measurement device actually does. The main point
was to extend results of Refs.~\cite{jackson2015,beck2017,jackson2017,jackson2017b,beck2018}, which were meant to provide tests for small correlated errors in quantum computing devices. While debugging small quantum computers forms one application of our scheme, 
it fits measurement device independent cryptography \cite{lo2012,twin2018,twin2019a,twin2019b,twin2019c} very well, and so we suggest our scheme can be fruitfully integrated with that protocol, too.

\section*{Acknowledgments}
We thank Mark Beck for useful discussions.

\bibliography{AliceTest2}
\end{document}